\documentclass[Journal,letterpaper,SectionNumbers,InsideFigs,NoLineNumbers]{ascelike-new}

\usepackage[utf8]{inputenc}
\usepackage{xcolor}
\usepackage[T1]{fontenc}
\usepackage{lmodern}
\usepackage{graphicx}
\usepackage[figurename=Fig.,labelfont=bf,labelsep=period]{caption}
\usepackage{subcaption}
\usepackage{amsmath}
\usepackage{adjustbox}
\usepackage{tikz}
\usepackage{enumitem}% http://ctan.org/pkg/enumitem
\usepackage{multirow}
\usepackage{booktabs}
\usepackage{comment}
\usepackage{multirow}
\usepackage{color}
\usepackage{colortbl}
\usepackage[normalem]{ulem}

\usepackage{newtxtext,newtxmath}
\usepackage[colorlinks=true,citecolor=black,linkcolor=black]{hyperref}

\newcommand{\possessivecite}[1]{\citeAN{#1}'s \citeyear{#1}}

\hypersetup{
    colorlinks = true,
    urlcolor   = black,
    citecolor  = black,
    linkcolor= black,
}

\NameTag{\small{Kramer et al. (2025), accepted in JHE}}
\DeclareMathOperator\erf{erf}
\usepackage{tikz,pgfplots}
\usepgfplotslibrary{groupplots}
\usepackage{pgfplotstable}
\usepgfplotslibrary{external}
\usetikzlibrary{positioning,arrows.meta,shapes}
\pgfplotsset{compat=1.15}

\usepackage{lipsum}
\usepackage{makecell}
\usepackage{longtable}

\begin{document}

% You will need to make the title all-caps

\title{Discussion on ``Resurrecting a Neglected Measurement Technique for Air–Water Flows''}

\author[1]{M. Kramer}
\author[2]{H. Wang}
\author[3]{D. B. Bung}

\affil[1]{Senior Lecturer, UNSW Canberra, School of Engineering and Technology (SET), Canberra,
ACT 2610, Australia,  ORCID 0000-0001-5673-2751, Email: m.kramer@unsw.edu.au}

\affil[2]{Professor, State Key Laboratory of Hydraulics and Mountain River Engineering, Sichuan University, Chengdu 610065, China, ORCID: 0000-0002-3542-8416, Email: hang.wang@uqconnect.edu.au}

\affil[3]{Professor, Hydraulic Engineering Section (HES), FH Aachen, University of Applied Sciences, Aachen 52066, Germany, ORCID 0000-0001-8057-1193, Email: bung@fh-aachen.de}

\maketitle
Over the last years, there has been a renewed interest in differentiating various contributions to the air concentration in high Froude-number self-aerated flows, see for example \citeN{Kramer2024}, comprising entrained and entrapped air. The former is characterised by entrained air packets and bubbles, while entrapped air corresponds to air transported along wave peaks and troughs. Entrapped air was first measured  by \citeN{Killen1968} using a so-called dipping probe, while a physical interpretation of the dipping probe signals was provided only later by \citeN{wilhelms2005bubbles}. 

Since then, it has been commonly accepted that two different measurement instruments, for example a dipping probe and a common phase-detection probe, are required to fully quantify entrained and entrapped air. Recently, an article entitled ``Resurrecting a Neglected Measurement Technique for Air–Water Flows'' was published by Wilhelms and Gulliver (2024, \url{https://doi.org/10.1061/JHEND8.HYENG-13904}), who re-iterated the importance of applying these concepts for cavitation prevention and air–water gas transfer, as well as the need for two separate measurement instruments. The authors are congratulated for their seminal works on entrained and entrapped air \cite{wilhelms2005bubbles,Wilhelms2024}, and it is stipulated that these concepts have been overlooked in the last two decades. 

In this discussion, a simple discrimination technique for phase-detection probe signals is proposed, which allows to differentiate entrained and entrapped air from existing datasets, recorded with a state-of-the-art dual-tip phase-detection probe. It is believed that this novel signal processing method will make \possessivecite{Killen1968}  dipping probe redundant, and that it will be useful for the validation of non-intrusive measurements of entrapped air, as well as for the development of  physics-based models for air-water mass transfer in self-aerated flows.  

\setcounter{section}{0}
\section{Considerations on dipping probe}
Let us revisit \possessivecite{Killen1968} dipping probe, which consisted of two electrodes, a surface electrode and an electrode placed at the channel bottom. The surface electrode had dimensions of 0.8 mm, 0.8 mm, and 6.4 mm in streamwise ($x$), vertical ($y$), and transverse ($z$) directions, and dipped in and out of the surface roughness as flow passed the probe. When the surface probe contacted water that had a continuous electrical path to the electrode at the channel bottom, a constant current was generated, which was translated into the fraction of air contained within the surface roughness, i.e., entrapped air \cite{wilhelms2005bubbles}.

It is important to note that the dipping probe relies on the same operational principle as state-of-the-art phase-detection probes, but there are two key differences: i) the two electrodes are spatially separated, and ii) the surface electrode has relatively large dimensions when compared to tip diameters of state-of-the-art phase-detection probes, with typical inner electrode diameters $\approx 0.1$ to 0.2 mm 
\cite[Table 1]{Kramer2024a}. The spatial separation implies that the dipping probe is not capable of detecting ejected water droplets \cite{Wilhelms2024}, while the larger size makes it insensitive to smaller air bubbles, which is corroborated by \possessivecite{Killen1968} description, stating that ``\textit{current response begins when the leading edge of the electrode enters the water surface and ends when the trailing edge leaves the water surface}''. It is known that the effect of ejected droplets on measured air concentrations is only minor \cite{Wilhelms2024}, and as such, the insensitivity of the dipping probe to smaller air  bubbles enabled the measurement of entrapped air. These considerations have important implications for the interpretation of previous results, and it is emphasized that 
\possessivecite{Killen1968} entrapped air concentrations are likely a function of the surface electrode's dimensions.

\section{Discrimination technique}
Moving on to state-of-the-art conductivity phase-detection probes, their needle tips consists of two closely spaced electrodes, which are referred to as inner and outer electrodes, see figures in \citeN{Kramer2024a}. Differences in conductivity of air and water allow for measuring the time that the tip is surrounded by air, denoted as chord times $t_{\text{ch},a}$ (in s), where the index $a$ stands for air. The time-averaged volumetric concentration of total conveyed air $\overline{c}$ can be written as 
\begin{equation}
\overline{c}  = \frac{\sum t_{\text{ch},a}}{T},
\label{eqCtot}
\end{equation}
where $T$ (in s) is the measurement duration. A simple discrimination technique is now formulated as 
\begin{equation}
\overline{c}_\text{ent}  = \frac{\sum t_{\text{ch,ent}}}{T},
\label{eqCent}
\end{equation}
\begin{equation}
\overline{c}_\text{trap}  = \frac{\sum t_{\text{ch,trap}} }{T},
\label{eqCtrap}
\end{equation}
where $\overline{c}_\text{ent}$ is the time-averaged concentration of entrained air, $\overline{c}_\text{trap}$ is the time-averaged concentration of entrapped air, $t_{\text{ch,ent}} = t_{\text{ch},a} \leq \mathcal{T}_\text{thres}$ is the chord time of entrained air (in s), $t_{\text{ch,trap}} = t_{\text{ch},a} > \mathcal{T}_\text{thres}$ is the chord time of entrapped air (in s), and $\mathcal{T}_\text{thres}$  (in s) is a threshold value, which can be related to an air bubble chord size $d_\text{thres}$ (in m) using
\begin{equation}
\mathcal{T}_\text{thres} = \frac{d_\text{thres}}{\overline{u}},
\label{eq:d}
\end{equation}
where $\overline{u}$ is the time-averaged streamwise interfacial velocity (in m/s). It is noted that the use of a single threshold assumes that the chord time of entrained bubbles is smaller than the chord time of entrapped air, which is anticipated to hold true for most cases. In order to apply the proposed discrimination technique, one has to select a threshold chord size $d_\text{thres}$, and this selection requires some additional considerations. 
In \possessivecite{Killen1968} measurements, $d_\text{thres}$ is believed to have corresponded to the largest dimension of the surface electrode ($d_\text{thresh} \approx 6.4$ mm), while specific details on the electrode's electrical response to the impact of various bubble sizes or clusters are unknown. A physics-based approach would relate $d_\text{thres}$ to a flow length scale, such as the Hinze diameter $d_h$ (in m), representing the limiting diameter where turbulence is no longer intense enough to break up an entrained air packets into smaller bubbles \cite{10.1007/978-981-97-4076-5_19}
\begin{equation}
d_h = \left( \frac{We_{c} \, \sigma}{2 \rho_w}  \right)^{3/5} \epsilon_w^{-2/5},
\end{equation}
where $\sigma$ (in N/m) is the surface tension coefficient between air and water, $We_\text{c}$ is the critical Weber-number, $\rho_w$ (in kg/m$^3$) is the density of water, and $\epsilon_w$ (in m$^2$/s$^3$) is the turbulent kinetic energy dissipation rate of water. In this context, it is stressed that i) critical Weber numbers have not been established for high-Froude number self-aerated flows, ii) an equilibrium state may not be present next to the air-water surface, and iii) the threshold $d_\text{thresh}$ may vary along the air-water flow column. In the absence of such knowledge, three constant threshold values of $d_\text{thresh} = 5$ mm, 10 mm, and 15 mm are adopted in the following, which is consistent with \possessivecite{Killen1968} surface probe dimension, as well as with recent optical measurements by \citeN{Wei2019} and \citeN{Wei2020}, who observed bubble sizes up to 10 mm in flows down a chute with smooth bed. Note that the mean bubble size is typically larger than the mean chord length, and, amongst others, a theoretical value of 1.5 has been reported for the ratio of bubble size to chord length \cite{LIU19931061,RUDISULI20121}, implying that  10 mm bubbles would correspond to $d_\text{thresh} = 15$ mm. 

\begin{figure}[h!]
\begin{center}
\includegraphics[angle=0]{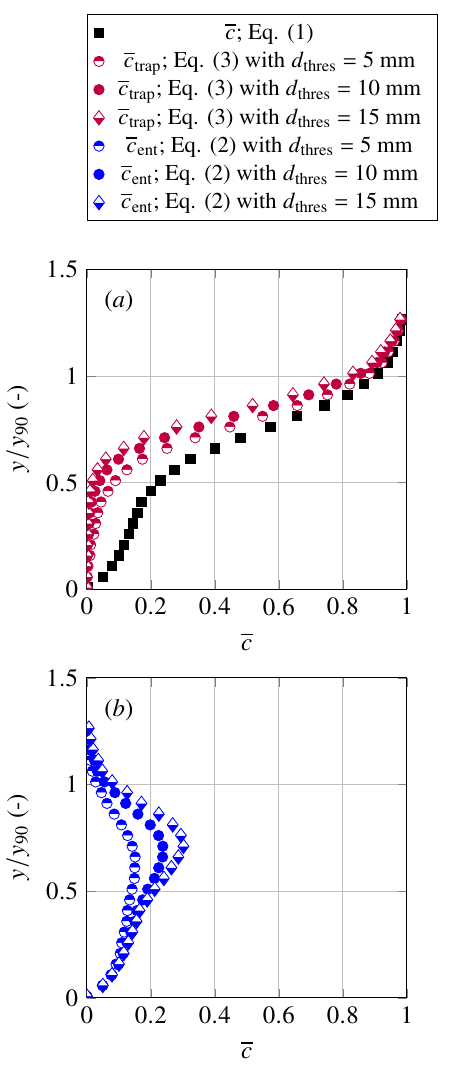}
\end{center}
\caption{Application of the discrimination technique to phase-detection probe signals of skimming flows down a stepped spillway with $d_c/h = 1.1$, chute angle $\theta = 45 ^\circ$, step edge = 11: (\textit{a})   total conveyed air $\overline{c}$ and entrapped air $\overline{c}_\text{trap}$; (\textit{b}) entrained air $\overline{c}_\text{ent}$}
\label{fig1}
\end{figure}

Here, the discrimination technique is applied to an experimental data set of skimming flows, recorded at a relatively large-sized stepped spillway model at the University of Queensland \cite{Kramer2018Transiton}. Phase-detection probe measurements were performed at the eleventh step edge for $d_c/h = 1.1$, where $d_c$ is the critical flow depth (in m), and $h = 0.1$ m is the step length. The chute angle was $\theta = 45 ^\circ$, and more details related to experimental facilities, flow measurement instrumentation, etc., are provided in \citeN{Kramer2018Transiton}. Figure \ref{fig1}\textit{a} shows distributions of total conveyed air ($\overline{c}$) and entrapped air ($\overline{c}_\text{trap}$), where the latter was evaluated with Eq. (\ref{eqCtrap}), using $d_\text{thres} = 5$ mm, 10 mm, and 15 mm, respectively. The ordinate is normalised with the mixture flow depth $y_{90} = y(\overline{c} = 0.9)$, and it is seen that the obtained distributions resemble the profile shapes of \citeN[pp 65 to 84]{Killen1968}, demonstrating  proof of concept of the proposed discrimination technique. Corresponding entrained air concentration distributions were evaluated using Eq. (\ref{eqCent}), and are shown in Fig. \ref{fig1}\textit{b}. It is discussed in the next section how these measurements can be reconciled with recently developed surface theories for mass diffusion in self-aerated flows.

\section{Modelling air concentration distributions}
The modelling of entrained and entrapped air concentration distributions in self-aerated flows has recently been described by \citeN{Kramer2023} and \citeN{Kramer2024}, who  characterised vertical mass diffusion using two flow momentum layers, namely a Turbulent Wavy Layer (TWL) and a Turbulent Boundary Layer (TBL). The entrapped and entrained air concentrations of the TWL are defined as \cite{Kramer2024}

\begin{equation}
\overline{c}_\text{trap}
= \frac{1}{2} \left(1 + \erf  \left( \frac{y - y_{50_\text{trap}}}{\sqrt{2} \mathcal{H}_\text{trap}} \right)  \right),
\label{entrappedair}
\end{equation}

\begin{eqnarray}
\overline{c}_\text{ent,TWL}= \frac{1}{2}  \left( \erf  \left( \frac{y - y_{50}}{\sqrt{2} \mathcal{H}} \right)  -  \erf  \left( \frac{y - y_{50_\text{trap}}}{ \sqrt{2} \, \mathcal{H}_\text{trap} } \right) \right),  \label{entrainedair}
\end{eqnarray}
where $y_{50_\text{trap}}$ (in m) is the free-surface level, $\mathcal{H}_\text{trap}$ (in m) is the root-mean-square wave height,  $y_{50} = y(\overline{c}=0.5)$ is the characteristic mixture flow depth (in m), and $\mathcal{H}$ (in m) is a characteristic length scale proportional to the thickness of the wavy surface-layer. Local superposition of $\overline{c}_\text{ent}$ and $\overline{c}_\text{trap}$ gives the air concentration of the TWL
\begin{equation}
\overline{c}_\text{TWL} = \overline{c}_\text{trap} + \overline{c}_\text{ent,TWL}.
\label{eqsuper}
\end{equation}

\begin{figure}[h!]
\begin{center}
\includegraphics[angle=0]{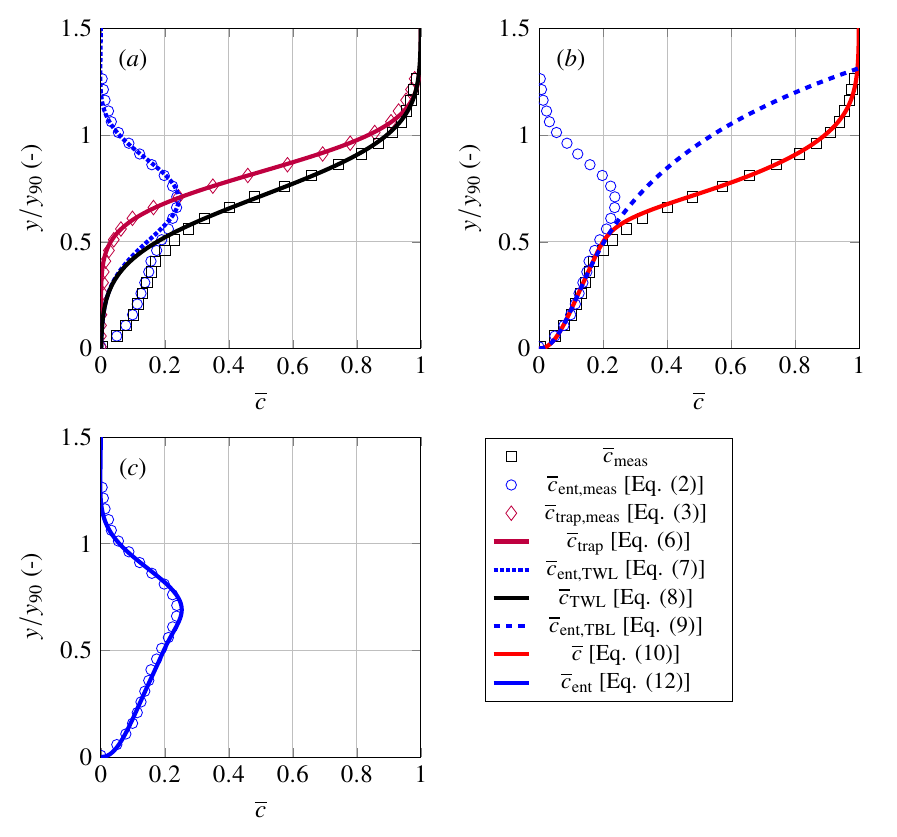}
\end{center}
\caption{Modelling of air concentration distributions using the combined two-sate convolution and superposition approach; $d_\text{thres} = 10$ mm: (\textit{a}) air concentration decomposition of the TWL;  (\textit{b}) modelled air concentration distributions of the TBL and total conveyed air; (\textit{c}) combined entrained
air concentration of the TWL and the TBL}
\label{fig2}
\end{figure}

Figure \ref{fig2}\textit{a} shows the application of Eqns. (\ref{entrappedair}) to (\ref{eqsuper}) to the re-analysed data set, where entrained air and entrapped air were evaluated using the discrimination technique, i.e., Eqns. (\ref{eqCent}) and (\ref{eqCtrap}), with $d_\text{thres} = 10$ mm. It is seen that the measured  profile of entrapped air shows excellent agreement with its analytical solution, and that  entrained air follows a Gaussian distribution within the TWL. Notably, around $y/y_{90} \lessapprox 0.5$, entrained air measurements start to deviate from Eq. (\ref{entrainedair}) (Fig. \ref{fig2}\textit{a}), which is because air bubbles are transitioning from the TWL into the TBL.

These entrained air bubbles are subsequently governed by different flow physics, described by an analytical solution of the advection diffusion equation for air in water \cite{Kramer2023}

\begin{equation}
\overline{c}_\text{ent,TBL} = 
\begin{cases}
\overline{c}_{\delta/2}  \left(\frac{y}{\delta-y} \right)^{\beta}, & y \leq \delta/2,
\label{eq:voidfraction1}\\
\vphantom{\left(\frac{\frac{\delta}{y_\star}-1}{\frac{\delta}{y}-1} \right)^{\frac{\overline{v}_r S_c}{\kappa u_*}}}
\overline{c}_{\delta/2} \, \exp \left(\frac{4\beta}{\delta} \left(y - \frac{\delta}{2} \right)   \right), \quad & y > \delta/2, \\
\end{cases}
\end{equation}
where $\overline{c}_{\delta/2}$ is the air concentration at half the boundary layer thickness (in m), and $\beta = \frac{\overline{v}_r S_t}{ 
 \kappa u_*}$ is the Rouse number for air bubbles in water, with $\overline{v}_r$ (in m/s) being the bubble rise velocity, $S_t$  the turbulent Schmidt number, $u_*$ (in m/s) the friction velocity, and $\kappa$ the von Karman constant. Equation (\ref{eq:voidfraction1}) is plotted together with entrained air measurements in Fig. \ref{fig2}\textit{b}, and the agreement between measurements and the analytical solution for $y/y_{90} \lessapprox 0.5$ confirms that the proposed discrimination technique allows to finely discern the flow physics of the different momentum layers. Next, a convolution of the TBL and the TWL state with a Gaussian interface probability led to the following expression for total conveyed air concentration \cite{Kramer2023}
\begin{equation}
\overline{c} = \left(\overline{c}_\text{trap} + \overline{c}_{\text{ent,TWL}} \right) \Gamma +  \overline{c}_{\text{ent,TBL}}(1-\Gamma), 
\label{eq:voidfractionfinal1}
\end{equation}
with
\begin{equation}
\Gamma = \frac{1}{2} \left( 1+\erf \left(\frac{y - y_\star }{ \sqrt{2} \sigma_\star}  \right)  \right).
\label{eq:gaussianerr}
\end{equation}
where  $\erf$ is the Gaussian error function, $y$ (in m) is the bed-normal coordinate, $y_\star$ (in m) is the time-averaged interface position, and $\sigma_\star$ (in m) is its standard deviation; note that the determination of all model parameters is discussed in \citeN{Kramer2023}. To resemble the combined entrained air concentration of the TWL and the TBL, Eq. (\ref{eq:voidfractionfinal1}) can be simplified to
\begin{equation}
\overline{c}_\text{ent} =  \overline{c}_{\text{ent,TWL}}  \Gamma +  \overline{c}_{\text{ent,TBL}}(1-\Gamma),
\label{eq:centcombined}
\end{equation}
where $\overline{c}_\text{trap}$ was neglected. Figures \ref{fig2}\textit{b,c} show the modelled total conveyed air and combined entrained air concentration distributions, demonstrating good agreement between the two-state convolution approach and measurements. Note that such agreement is somewhat expected, as Eq. (\ref{eq:voidfractionfinal1}) has previously  been validated with more than 500 data sets from literature \cite{Kramer2023}. Lastly, it is stressed that the presented conceptualisation allows to distinguish three different physical mechanisms contribution to the air concentration, see Eq. (\ref{eq:voidfractionfinal1}), comprising 
entrapped air within the TWL, entrained air within the TWL, and entrained air within the TBL \cite{Kramer2024}, and that the presented discrimination technique allows to unravel these contributions.

\section{Conclusion}
In this short discussion, a novel discrimination technique for the analysis of phase-detection probe signals is introduced, which enables the differentiation of entrained and entrapped air in high Froude-number self-aerated flows. Importantly, the discrimination technique only requires one state-of-the-art phase-detection probe, thereby rendering the need for \possessivecite{Killen1968} dipping probe.
The implementation of the discrimination technique is relatively simple and straightforward, while it is acknowledged that the threshold selection is not trivial. Ideally, the threshold should be related to a characteristic time or length scale of the flow, such as the Hinze scale, and further research is warranted. Noting that the dimensions of \possessivecite{Killen1968} dipping probe are fixed, the discrimination technique also offers more flexibility in performing such analyses. Overall, it is anticipated that the contribution of this discussion will be useful for future validation of non-intrusive measurements of air water flow properties, as well as for the development of more sophisticated air-water mass transfer models. 

\section*{Data Availability Statement}
No data, models, or code were generated or used during the study.

\section*{Acknowledgments}
Daniel Valero  is thanked for fruitful discussions. 

\bibliography{ascexmpl-new}

\end{document}